\documentclass[12pt]{article}
\usepackage{amssymb}
\usepackage{epsfig,epsf,graphicx}
\usepackage{amsmath}
\def\be{\begin{eqnarray} &&}

\def\ee{\end{eqnarray}}

\def\bew{\begin{widetext}}
\def\ew{\end{widetext}}

\usepackage[usenames]{color}
\usepackage[normalem]{ulem}

\begin{document}

\begin{titlepage}

\begin{center}
{\Large\bf{Excited states of the Wick-Cutkosky model with the Nakanishi representation in the Light-Front framework}}
\end{center}

\begin{center}
\bf{R. Pimentel and W. de Paula}\\
{\sl Dep. de F\'\i sica, Instituto Tecnol\'ogico de Aeron\'autica, CTA, 12228-900 \\ S\~ao Jos\'e dos Campos, Brazil}
\end{center}

\begin{abstract}
The Wick-Cutkosky model is investigated using the framework of the Nakanishi Perturbative Integral Representation projected in the Light-Front hyperplane for an s-wave amplitude. We developed a new technique based on sucessive integration by parts of the Nakanishi Representation which enabled us to transform one of the integrations into a sum. With a set of boundary conditions it was possible to guess the format of the solution, which was in fact a distribution. The eigenequations obtained were the same as the originals of Cutkosky \cite{CutPR54}. Finally, a numerical check confirmed that the final equation reproduce the same eigenvalues as the initial Bethe-Salpeter equation.
%
%\keywords{Nakanishi Representation \and Wick-Cutkosky Model \and Bethe-Salpeter Equation}
\end{abstract}

\end{titlepage}
%\institute{
%R. Pimentel \at
%              Instituto Tecnol\'ogico de Aeron\'autica, DCTA, 12.228-900, S\~ao Jos\'e dos Campos, Brazil\\
%              Tel.: +55-12-39475939\\
%              Fax: +55-12-3947\\
%              \email{pimentel.es@gmail.com}           %  \\
%\and
%W. de Paula \at
 %             Instituto Tecnol\'ogico de Aeron\'autica, DCTA, 12.228-900, S\~ao Jos\'e dos Campos, Brazil\\
%              Tel.: +55-12-39475939\\
%              Fax: +55-12-3947\\
%              \email{wayne@ita.br}           %  \\

%\maketitle

\section{Introduction}
\label{intro}
The Bethe-Salpeter (BS) equation \cite{SalPR51} was proposed to treat the problem of the bound state in a relativistic framework. However, since it's kernel had poles, the numerical solution could not be obtained using standard numerical algorithms for integral equations. In order to solve this difficulty, in $1954$ Wick \cite{WicPR54} and Cutkosky \cite{CutPR54} published articles proposing new methods. In particular, the Wick rotation avoids the poles and, since then, this has been the canonical way to solve the BS equation. Nevertheless, Wick also proposed the strategy of using a particular type of integral representation for the amplitude of the BS equation with a massless interaction, and using a Dirac delta Ansatz for his weight function he was able to obtain a simpler equation for the massless case. Moreover, Cutkosky extended Wick's work to the full spectrum of the massless ladder interaction case. This model has been known as the Wick-Cutkosky model, specially important as a toy model because of it's simplicity.

In the seventies, Nakanishi explored the idea of an integral representation, very similar to the type used by Wick, for Feynman Diagrams and scattering amplitudes \cite{NakPR63,NakPTP69,Nak71}

\begin{equation}
	\phi(k) = \int^{\infty}_{0} d \gamma' \int ^{1}_{0} \prod_{h} dz'_{h} \delta ( \sum_{h}z'_{h} -1)  \frac {g^{(n)}(\gamma',z')}{(\gamma'-\sum_{h} z'_h s_h -i\epsilon)^n},
	\label{Naka}
\end{equation}
where $s_h$ are scalar products of the external momenta. He analyzed this representation and demonstrated important properties such as analicity and uniqueness. Interestingly, although Nakanishi studied this formula for perturbative purposes, the initial ideia of Wick was to use it to solve the nonperturbative BS equation.

In $1995$ Kusaka and Williams observed that the Nakanishi representation was useful to solve the massive interaction BS equation in Minkowski space \cite{KusPRD95, KusPRD97}, but the method was still not very practical for numerical purposes. Thus, in $2006$ Karmanov and Carbonell proposed a new technique based on the application of the Nakanishi representation to the BS equation and then projecting it in the Light-Front hyperplane \cite{KarEPJA06,CarEPJA06,CarEPJA09,CarEPJA10}, and this resulted in an generalized integral equation with smooth kernels. Then Frederico, Salme and Viviani extended the Light Front projected BS equation for scattering states and also developed a new method based on the uniqueness of the Nakanishi representation \cite{FrePRD12,FreFBS11,FrePRD14}.

Following the recent success of the Light Front projected BS equation technique, we decided to look again to the massless interaction case, which was the first one described by an integral representation. We found that the solutions that before were thought of as an Ansatz could be obtained by using integration by parts and suitable boundary conditions, even for the excited states. Moreover we checked numerically the eigenvalues of the obtained equation and the original one.

%%%%%%%%%%%%%%%%%%%%%%%%%%%%%%%%%%%%%%%%%%%%%
\section{Wick-Cutkosky Model with the Nakanishi Representation}
The scalar BS equation with a ladder massless interaction is given by
\begin{equation}
\phi(k,p) = \frac{1}{m^2-(\frac{p}{2} -k)^2}\frac{1}{m^2-(\frac{p}{2}+k)^2} i g^2 \int \frac{d^4 k'}{(2\pi)^4} \frac{\phi(k',p)}{(k-k')^2+i\epsilon},
\end{equation}
where $ g$ is the coupling constant, $p$ is the total momentum of the bound state and $k$ is the relative momentum.

The BS amplitude $\phi(k,p)$  can be written as a three leg Nakanishi representation with one momentum on-shell ($p$) and two off-shell ($\frac{p}{2}-k$ and $\frac{p}{2}+k$),

\begin{equation}
	\phi(k,p) = \frac{-i}{4 \pi}\int^{\infty}_{0} d \gamma' \int ^{1}_{-1} dz' \frac {g(\gamma',z')}{(\gamma'+\kappa^2-k^2- p \cdot k z'-i\epsilon)^3}.
\label{NakaPhi}
\end{equation}

Where $\kappa^2 = m^2 - \frac{p^2}{4}$. Note that the symmetry $\phi(k_0,\vec{k},p) = \phi(-k_0,\vec{k},p)$, such as for an s-wave, is translated into $g(\gamma',z') = g(\gamma',-z') $ for the weight function in this representation.
Substituting this representation in the previous BS equation and projecting the equation in the Light Front by integrating in $k^{-}$ we obtain \cite{KarEPJA06}

\begin{equation}
\int_{0}^{\infty} d\gamma' \frac{g(\gamma', z) }{\left(\gamma'+\gamma+z^2m^2+(1-z^2)\kappa^2\right)^2} = \int_{0}^{\infty} d\gamma' \int_{-1}^{1} dz' ~V(\gamma, z; \gamma', z') g(\gamma',z'). \label{BSE}
\end{equation}
The Kernel of the integral equation is
\begin{small}
\begin{eqnarray}
&&V(\gamma, z; \gamma', z') = \frac{\alpha m^2}{2\pi} \frac{1}{\left(\gamma+z^2m^2+(1-z^2)\kappa^2\right)}\frac{1}{\left(\gamma'+z'^2m^2+(1-z'^2)\kappa^2\right)}\nonumber\\
&&\times\left(\frac{\theta(z-z')}{\left(\gamma + \gamma' \frac{1-z}{1-z'} + z^2m^2+(1-z^2)\kappa^2\right)}\frac{1-z}{1-z'} + \frac{\theta(z'-z)}{\left(\gamma + \gamma' \frac{1+z}{1+z'} + z^2m^2+(1-z^2)\kappa^2\right)}\frac{1+z}{1+z'}\right),\nonumber\\
\end{eqnarray}
\end{small}
where we defined $\alpha = g^2/4\pi$. The BS equation can be rewritten in a suitable way as
\begin{eqnarray}
&&\int_{0}^{\infty} d\gamma' \frac{g(\gamma', z) }{\left(\gamma'+\gamma+z^2m^2+(1-z^2)\kappa^2\right)^2} = \nonumber\\
&&\frac{\alpha m^2}{2\pi} \frac{1}{d_0(z)} \int_{-1}^{1} dz' \left[\omega(z,z')\theta(z-z')+\omega(-z,-z')\theta(z'-z)\right],
\end{eqnarray}
where we introduced the auxiliary functions
\begin{eqnarray}
\omega(z,z') &=& \int_{0}^{\infty} d\gamma' \frac{g(\gamma', z') }{(\gamma'+a(z'))(\gamma'+c(\gamma,z,z'))};~~~ d_{0}(\gamma,z) = \gamma+z^2m^2+(1-z^2)\kappa^2;\nonumber\\
a(z') &=& z'^2m^2+(1-z'^2)\kappa^2; ~~~~ c(\gamma,z,z')=\frac{1-z'}{1-z}d_{0}(\gamma,z).
\end{eqnarray}
Now, we intend to transform the integration in $\gamma'$ into a sum, using integration by parts. To do that, we first define the sequence $G^{(n)}\left(\gamma',z\right)$ as
\begin{eqnarray}
G^{(0)}(\gamma',z) &=& g(\gamma',z); ~~~~G^{(n+1)}(\gamma',z) = - \int_{\gamma'}^{\infty} d\gamma'' ~G^{(n)}(\gamma'',z);\nonumber\\
&&G^{(n)}(\gamma',z) = \frac{\partial}{\partial\gamma'} G^{(n+1)}(\gamma',z).
\end{eqnarray}
First, we perform an integration by parts in $\gamma'$ in the LHS of Eq. \eqref{BSE}
\begin{eqnarray}
&&\int_{0}^{\infty} d\gamma'\frac{g(\gamma', z) }{\left(\gamma'+\gamma+z^2m^2+(1-z^2)\kappa^2\right)^2} = \int_{0}^{\infty} d\gamma' \frac{g(\gamma', z) }{\left(\gamma' + d_{0}(\gamma,z)\right)^2} =\nonumber\\
&& = \int_{0}^{\infty} d\gamma' \left(\frac{\partial}{\partial\gamma'}G^{(1)}(\gamma', z) \right) \frac{1}{\left(\gamma' + d_{0}(\gamma,z)\right)^2} = \nonumber\\
&=&  G^{(1)}(0, z)  \frac{1}{d_{0}(\gamma,z)^2} -
\int_{0}^{\infty} d\gamma' ~G^{(1)}(\gamma',z)\frac{\partial}{\partial\gamma'}\left(\frac{1}{\left(\gamma' + d_{0}(\gamma,z)\right)^2}\right). \nonumber\\
\end{eqnarray}

Each integration by parts generates a new term for the sum. In general, the relationship that generates the n-th term of the sum is
\begin{small}
\begin{eqnarray}
&&\int_{0}^{\infty} d\gamma' ~G^{(n)}(\gamma',z)\frac{\partial^{(n)}}{\partial\gamma'^{(n)}} \left(\frac{1}{\left(\gamma' + d_{0}(\gamma,z)\right)^2}\right) = \nonumber\\
&=&\int_{0}^{\infty} d\gamma' \left(\frac{\partial}{\partial\gamma'} G^{(n+1)}(\gamma',z)\right) \frac{\partial^{(n)}}{\partial\gamma'^{(n)}} \frac{1}{\left(\gamma' + d_{0}(\gamma,z)\right)^2} =\nonumber\\
&=&\left(G^{(n+1)}(\gamma',z) \frac{\partial^{(n)}}{\partial\gamma'^{(n)}} \left(\frac{1}{\left(\gamma' + d_{0}(\gamma,z)\right)^2}\right)\right){\bigg|}_{0}^{\infty} \nonumber\\
&-&\int_{0}^{\infty} d\gamma' ~ G^{(n+1)}(\gamma',z) \frac{\partial^{(n+1)}}{\partial\gamma'^{(n+1)}} \left(\frac{1}{\left(\gamma' + d_{0}(\gamma,z)\right)^2}\right).\nonumber\\
\end{eqnarray}
\end{small}
Thus, after performing $n$ integrations by parts we have

\begin{eqnarray}
&&\int_{0}^{\infty} d\gamma' \frac{g(\gamma', z) }{\left(\gamma'+\gamma+z^2m^2+(1-z^2)\kappa^2\right)^2} = \nonumber\\
&=& \sum_{i=1}^{n}(-1)^{i-1}\left(G^{(i)}(\gamma',z) \frac{\partial^{(i-1)}}{\partial\gamma'^{(i-1)}} \left(\frac{1}{\left(\gamma' + d_{0}(\gamma,z)\right)^2}\right)\right){\bigg |}_{0}^{\infty} \nonumber\\
&+& (-1)^{n}
\int_{0}^{\infty} d\gamma' ~ G^{(n)}(\gamma',z) \frac{\partial^{(n)}}{\partial\gamma'^{(n)}} \left(\frac{1}{\left(\gamma' + d_{0}(\gamma,z)\right)^2}\right).\nonumber\\
\end{eqnarray}

The term in the limit of large $n$ vanishes and each term of the sum is proportional to $ G^{(i)}(0,z) $. So, we define this term as a sequence of functions in the variable z

\begin{equation}
b_{i}(z) = G^{(i)}(0,z)
\end{equation}

Finally, in the limit of $n\rightarrow\infty$

\begin{eqnarray}
\label{LHSsum}
&&\int_{0}^{\infty} d\gamma' \frac{g(\gamma', z) }{\left(\gamma'+\gamma+z^2m^2+(1-z^2)\kappa^2\right)^2}=
\sum_{i=1}^{\infty} i!\frac{b_{i}(z)}{d_{0}(\gamma,z)^{i+1}}.\nonumber\\
\end{eqnarray}

Note that we successfully rewrote the integration in $\gamma'$ as a sum. We now intend to do the same for the RHS of Eq. \eqref{BSE}. In the following we give some details of the derivation.
\begin{small}
\begin{eqnarray}
\label{RHSsum}
&&\int_{0}^{\infty} d\gamma' \frac{g(\gamma', z') }{(\gamma'+a(z'))(\gamma'+c(\gamma,z,z'))}=\nonumber\\
&=&\sum_{j=1}^{\infty}(-1)^{j-1} \left(G^{(j)}(\gamma',z') \frac{\partial^{(j-1)}}{\partial\gamma'^{(j-1)}} \left(\frac{1}{\left(\gamma' + a(z')\right)\left(\gamma' + c(\gamma,z,z')\right)}\right)\right){\bigg|}_{0}^{\infty}\nonumber\\
&=&\sum_{j=1}^{\infty}(-1)^{j-1} b_j(z') \sum_{n=0}^{j-1} \binom{j-1}{n}  \frac{\partial^{(j-n-1)}}{\partial\gamma'^{(j-n-1)}}
\left(\frac{1}{\gamma' + a(z')}\right)
\frac{\partial^{(n)}}{\partial\gamma'^{(n)}}\left(\frac{1}{\gamma' + c(\gamma,z,z')}\right)\nonumber\\
&=&\sum_{j=1}^{\infty} b_j(z') \sum_{n=0}^{j-1}  \frac{(j-1)!}{n!(j-n-1)!} \frac{(j-n-1)!}{a(z')^{j-n}}\frac{n!}{c(\gamma,z,z')^{n+1}}\nonumber\\
&=&\sum_{j=1}^{\infty} b_j(z') \sum_{n=0}^{j-1}  \frac{(j-1)!}{a(z')^{j-n} c(\gamma,z,z')^{n+1}}\nonumber\\
&=&\sum_{i=1}^{\infty} \frac{1}{c(\gamma,z,z')^i} \sum_{j=i}^{\infty}  \frac{(j-1)! b_j (z')}{a(z')^{j-i+1} }\nonumber\\
&=&\sum_{i=1}^{\infty} \left( \frac{1-z}{(1-z')d_0(z)} \right)^i \sum_{j=i}^{\infty}  \frac{(j-1)! b_j (z')}{a(z')^{j-i+1} }
\end{eqnarray}
\end{small}
Collecting the terms, the BS equation for the Wick-Cutkosky model can be expressed as
\begin{small}
\begin{eqnarray}
\sum_{i=1}^{\infty} i!\frac{b_{i}(z)}{d_{0}(\gamma,z)^{i+1}} &=& \frac{\alpha m^2}{2\pi} \frac{1}{d_0(z)} \int_{-1}^{1} dz' \sum_{i=1}^{\infty} \left( \frac{1-z}{(1-z')d_0(z)} \right)^i \sum_{j=i}^{\infty}  \frac{(j-1)! b_j (z')}{a(z')^{j-i+1} } \theta(z-z') \nonumber\\
 &+& z\rightarrow -z ~ \text{and} ~z' \rightarrow -z'
\end{eqnarray}
\end{small}
Note that for $ \gamma \rightarrow \infty \Rightarrow d_0 \rightarrow \infty $ and therefore the small $i$ terms dominates the series. Then we can match the $\frac{1}{d_0^i}$ terms of the series and we have, since $a(z)$ and $b_i(z)$ are even:
\begin{small}
\begin{equation}
b_i(z) = \frac{\alpha m^2}{2 \pi} \int_{-1}^{1} dz' \frac{1}{i!} \left( \left( \frac{1-z}{1-z'} \right)^i \theta(z-z') + \left( \frac{1+z}{1+z'} \right)^i \theta(z'-z)\right) \sum_{j=i}^{\infty}  \frac{(j-1)! b_j (z')}{a(z')^{j-i+1} }.
\end{equation}
\end{small}
To solve the eigenequation we set $b_i(z) = 0$ if $i > i_{max}$ for some $i_{max}$. Note that this is consistent because all the equations for $i > i_{max}$ becomes trivially solved, since they are homogeneous and each $b_i$ depends only on $b_j$ with $j>i$. Thus, it is only necessary to solve the equations with $i \leqslant i_{max}$.  Since the resulting matrix is triangular by blocks, the eigenvalues are solely determined by the $i=j=i_{max}$ equations:
\begin{equation}
\label{wickcut}
b_i(z) = \frac{\alpha m^2}{2 \pi} \int_{-1}^{1} dz' \frac{1}{i} \left( \left( \frac{1-z}{1-z'} \right)^i \theta(z-z') + \left( \frac{1+z}{1+z'} \right)^i \theta(z'-z) \right)  \frac{ b_i (z')}{a(z') },
\end{equation}
which is exactly the Wick-Cutkosky equation originally obtained. It is known that the support for the Nakanishi weight function for the Wick-Cutkosky model is only $\gamma'=0$. Previously it was used the Ansatz $g(\gamma',z')=\delta(\gamma')f(z')$ which is equivalent to set $b_i(z') = 0 $ for $ i > 1$ in the present formalism. To fully reconstruct $g(\gamma',z')$ from the $b_i(z')$ we propose the following Ansatz:
\begin{equation}
g(\gamma',z') = \sum_{i=1}^{i_{max}} b_i(z')\delta^{(i-1)}(\gamma').
\label{weight}
\end{equation}
This distribution has support only in $\gamma' = 0$ and clearly reproduces equations \eqref{LHSsum} and \eqref{RHSsum}. This happens because the derivative of delta acts on the kernel in the same way as when one performs the integration by parts , thus this is the solution for the Nakanishi weight function. At this point we notice that the integration by parts is ill-defined since the weight function is a distribution and not a function. To be mathematically precise, first one should set a small mass $\mu$ for the interaction particle and then take the limit of $\mu \rightarrow 0$ in the end of the calculation.
To reconstruct the BS amplitude we can use the Nakanishi Perturbative Integral Representation \eqref{NakaPhi}.
%\begin{equation}
%	\phi(k,p) = \frac{-i}{4 \pi}\int^{\infty}_{0} d \gamma' \int ^{1}_{-1} dz' \frac {g(\gamma',z')}{(\gamma'+\kappa^2-k^2-p.kz'-i\epsilon)^3}.
%	\label{Naka}
%\end{equation}
Using the weight function \eqref{weight} in the representation \eqref{NakaPhi} we have
\begin{equation}
	\phi(k,p) = \frac{-i}{4 \pi} \sum^{i_{max}}_{n=1} \frac{(n+1)!}{3} \int ^{1}_{-1} dz' \frac {b_{n}(z')}{(\kappa^2-k^2-p \cdot k z'-i\epsilon)^{n+2}}.
	\label{BSA}
\end{equation}
From \eqref{BSA} we can also obtain the light-front wave function using the relation
\begin{equation}
	\psi(\bold{k_\perp},z) = \frac{(\omega.k_1)(\omega.k_2)}{\pi (\omega.p)} \int ^{\infty}_{- \infty}  \phi (k+\beta \omega,p)d\beta,
	\label{LFProj}
\end{equation}
where $\omega = (1,0,0,-1)$. Finally, we have
\begin{equation}
	\psi(\bold{k}_\perp,z) = \frac{1-z^2}{8 \sqrt{\pi}} \sum^{i_{max}}_{i=1} i! \frac{b_i(z)}{(\bold{k}^2_\perp + m^2-(1-z^2)\frac{M^2}{4})^{i+1}}.
	\label{LFWF}
\end{equation}

\section{Numerical Results}
\label{numerical}
In this section we compare the numerical values of the first four eigenvalues by solving the Bethe-Salpeter equation for zero mass exchange,  Eq. \eqref{BSE}, and the ones obtained from Eq. \eqref{wickcut}, which is obtained using the method described in the last section. \\
We solve the BS equation, Eq. \eqref{BSE}, using a Laguerre basis in $\gamma$ and $\gamma'$, and a Legendre basis in $z$ and $z'$. Moreover, with 5 Laguerre basis functions and 5 Legendre polynomials, we obtain the eigenvalues of the ground state ($\alpha_1^{(\gamma',z')}$) and the first excited state ($\alpha_2^{(\gamma',z')}$) for various bounding energies $B/m$ .
Also, we solve Eq. \eqref{wickcut} using a Legendre basis in z and $z'$, obtaining the eigenvalues of the ground state ($\alpha_1^{(z')}$) and the first excited state ($\alpha_2^{(z')}$). The results are in Table \ref{table:eigenvalues}.
\begin{table} [htb]
\caption
{Values of $\alpha=g^2/4\pi$ as a function of binding energy $B/m$ for a massless scalar exchange for the ground state and the first excited state. The first and third column shows the results for Eq.  \eqref{BSE}. The second and fourth column shows the results for Eq. \eqref{wickcut}.}
 \centering
\begin{tabular}{c c c c c}
 \hline \hline $B/m$ & $\alpha_1^{(\gamma',z')}$ & $\alpha_1^{(z')}$ & $\alpha_2^{(\gamma',z')}$ &  $\alpha_2^{(z')}$   \\
 \hline
 0.1 & 1.11 & 1.12 & 2.90 & 2.93  \\
 0.2 & 1.78 & 1.79 & 4.90 & 4.85  \\
 0.3 & 2.34 & 2.35 & 6.58 & 6.53  \\
 0.4 & 2.84 & 2.84 & 8.09 & 8.05  \\
 0.5 & 3.29 & 3.30 & 9.44 & 9.42  \\
 \hline
\end{tabular}
\label {table:eigenvalues}
\end{table}\\
For the ground state and the first excited state the numerical results has a relative error of about $1\%$, which support the method presented in this work. \section{Summary}
\label{summary}
The Wick-Cutkosky model was reexamined using the new light front projected Nakanishi method.

In this case, the expansion in basis functions is not efficient, because the solution is a distribution. Thus, it was necessary to find an alternative way to deal with it. Our strategy was to transform the integration in $\gamma'$ into a sum, using integration by parts and a suitable boundary condition for the solution. This enabled us not only to reproduce the equations originally obtained by Cutkosky, but also to show that the weight-function for the amplitude reduces to a distribution in the massless limit. So, we were able to use these weight-functions to write the final formulas for the Bethe-Salpeter amplitude and the light-front wave function.

Moreover, we checked the final and initial equations numerically, expanding the solution in Laguerre and Legendre basis, and we obtained the same eigenvalues for the ground state and for the first excited states in both equations.

The integration-by-part method may also be useful in the calculation of the BS equation with a small interacting mass, specially if used with the uniqueness equation. This is importante because the basis expansion performs better when the interacting mass is bigger and thus the weight function has a more distributed support in $\gamma'$.

Following this procedure, the next step is to obtain the eigenvalues of higher excited states and their corresponding wave functions, which is a work in progress.


\begin{thebibliography}{99}
\bibitem{CutPR54}  Cutkosky, R.E.: Solutions of a Bethe-Salpeter Equation. Phys. Rev. \textbf{96}, 1135 (1954).
\bibitem{SalPR51}  Salpeter, E.E., Bethe,H.A.: A Relativistic Equation for Bound-State Problems. Phys. Rev. \textbf{84}, 1232 (1951).
\bibitem{WicPR54}  Wick, G.C.: Properties of Bethe-Salpeter Wave Functions. Phys. Rev. \textbf{96}, 1124 (1954).
\bibitem{NakPR63}  Nakanishi, N.: Partial-Wave Bethe-Salpeter Equation. Phys. Rev. \textbf{130}, 1230 (1963).
\bibitem{NakPTP69} Nakanishi, N.: A General survey of the theory of the Bethe-Salpeter equation. Prog. Theor. Phys. Suppl. 43, 1 (1969).
\bibitem{Nak71}    Nakanishi, N.: Graph Theory and Feynman Integrals.	(Gordon and Breach, New York, 1971).
\bibitem{KusPRD95} Kusaka,K., Williams, A.G.: Solving the Bethe-Salpeter equation for scalar theories in Minkowski space.  Phys. Rev. {\bf D 51}, 7026 (1995)
\bibitem{KusPRD97} Kusaka, K., Simpson, K.,  Williams, A.G.: Solving the Bethe-Salpeter equation for bound states of scalar theories in Minkowski space. Phys. Rev. {\bf D 56}, 5071 (1997).
%\bibitem{WickPR54} G. C.~Wick,     Phys. Rev. {\bf 96}, 1124 (1954).
\bibitem{KarEPJA06} 		Karmanov, V.A., Carbonell, J.: Solving Bethe-Salpeter equation in Minkowski space. Eur. Phys. J. A \textbf{27}, 1 (2006).
%Solutions of Bethe-Salpeter and light-front equations with cross-ladder kernel
\bibitem{CarEPJA06} 		Carbonell, J.,  Karmanov, V.A.: Cross-ladder effects in Bethe-Salpeter and light-front equations.  Eur. Phys. J.  A \textbf{27}, 11 (2006)
%Electromagnetic form factor via Bethe-Salpeter amplitude in Minkowski space
\bibitem{CarEPJA09} 		Carbonell, J., Karmanov, V.A., Mangin-Brinet, M.: Electromagnetic form factor via Bethe-Salpeter amplitude in Minkowski space.  Eur. Phys. J. A39, 53 (2009)
%Solving Bethe-Salpeter equation for two fermions in Minkowski space
\bibitem{CarEPJA10}  Carbonell, J., Karmanov, V.A.: Solving Bethe-Salpeter equation for two fermions in Minkowski space.  Eur. Phys. J.  A46, 387 (2010)
\bibitem{FrePRD12} 			Frederico, T., Salm\`e, G., Viviani, M.: Two-body scattering states in Minkowski space and the Nakanishi integral representation onto the null plane. Phys. Rev. D \textbf{85}, 036009 (2012).
\bibitem{FreFBS11} 			Frederico, T., Salm\`e, G.: Projecting the Bethe-Salpeter Equation onto the Light-Front and Back: A Short Review. Few-body Syst. {\bf 49}, 163 (2011).
\bibitem{FrePRD14} 		Frederico, T., Salm\`e, G., Viviani, M.: Quantitative studies of the homogeneous Bethe-Salpeter Equation. Phys. Rev. D \textbf{89}, 016010 (2014).
%\bibitem{SalPRC00}			J. H. O. Sales, T. Frederico, B. V. Carlson, P.	U. Sauer, Phys. Rev. {\bf C 61}, 044003 (2000).
%Quantitative studies of the homogeneous Bethe-Salpeter equation in Minkowski space
%Bethe-Salpeter scattering amplitude in Minkowski space
%\bibitem{CarPLB13} J. Carbonell, V.A. Karmanov, Phys. Lett.  B727, 319 (2013)
%Solving Bethe-Salpeter Equation for Scattering States
%\bibitem{KarFBS13} V.A. Karmanov, J. Carbonell, Few Body Syst. 54, 1509 (2013)
\end{thebibliography}
\end{document}